# Efficient simulation of many-body localized systems

Yichen Huang (黄溢辰)*
*Department of Physics, University of California, Berkeley, Berkeley, California 94720, USA*
*Institute for Quantum Information and Matter, California Institute of Technology, Pasadena, California 91125, USA*
(Dated: August 19, 2015)

An efficient numerical method is developed using the matrix product formalism for computing the properties at finite energy densities in one-dimensional (1D) many-body localized (MBL) systems. Arguing that any efficient (possibly quantum) algorithm can only have a polynomially small energy resolution, we propose a (rigorous) polynomial-time (classical) algorithm that outputs a diagonal density operator supported on a microcanonical ensemble of an inverse polynomial bandwidth. The proof uses no other conditions for MBL but assumes that the effect of any local perturbation (e.g., injecting conserved charges) is restricted to a region whose radius grows logarithmically with time. A non-optimal version of this algorithm efficiently simulates the quantum phase estimation algorithm in 1D MBL systems; a heuristic version of the algorithm can be easily coded and used to, e.g., detect energy-tuned dynamical quantum phase transitions between MBL phases. We extend the algorithm to two and higher spatial dimensions using the projected entangled pair formalism.

Classical simulation of quantum many-body systems is a major area of research in condensed matter physics [39]. In general, quantum systems cannot be efficiently simulated classically for the simple reason that the dimension of the Hilbert space grows exponentially with the system size. Remarkably, most physical systems are highly non-generic in that they are local. Indeed, locality is essential to the design of efficient classical algorithms. At a basic level, any local Hamiltonian with short-range interactions satisfies the Lieb-Robinson bound, which states that correlations can only propagate at a constant speed ("linear light cone") [22, 31, 34]. Consequently, the short-time dynamics of 1D local Hamiltonians allows efficient classical simulation [38]. The locality of a quantum state can be characterized by (i) exponential decay of correlations, (ii) area law for entanglement [17], (iii) efficient tensor network representation [14, 37, 53], etc. For example, the ground states of 1D gapped [24] and critical [52] Hamiltonians are well approximated by matrix product states (MPS) [18, 42] of small bond dimension. As the leading numerical method for computing the ground states of 1D local Hamiltonians, the density matrix renormalization group (DMRG) [57, 58] is a variational algorithm over MPS.

MBL is an active area of research studying a class of interacting quantum systems that are localized in a much stronger sense [1, 35]. In the literature, there are several (inequivalent) characterizations of MBL: (I) absence of transport and vanishing dc conductivity [4, 20], (II) logarithmic growth of entanglement with time [3, 26, 48, 55, 56, 60], (III) area law for the entanglement of and efficient tensor network representation of (almost) all eigenstates [6, 12, 19, 47], (IV) (quasi-)local integrals of motion [13, 26, 28, 45, 47], etc. One might expect that MBL systems allow efficient classical simulation by making use of their locality. Indeed, (II) strongly suggests that the time-evolving block decimation (TEBD) algorithm [54] can efficiently simulate the long-time dynamics of 1D MBL systems, cf. [50] is able to work with up to 40 spin-1/2's. For static properties of MBL systems at finite energy densities, exact diagonalization is the only prevalent numerical method to date, but it is limited to small system sizes [30, 33, 36, 40], and finite-size effects are not always negligible, cf. the state-of-the-art exact diagonalization is able to work with Hilbert spaces of dimension 705432 or $\lesssim 20$ spin-1/2's [33]. It is highly desirable to develop efficient MPS-based algorithms for computing the properties of excited states in 1D MBL systems.

Ideally, one might wish to compute an exact eigenstate (up to the truncation of real numbers) rather than a superposition or mixture of eigenstates that are close in energy. This requires a resolution of the order of the level spacing in the energy spectrum, which is exponentially small in the system size. We believe that any (possibly quantum) algorithm for our purposes satisfies the "time-energy uncertainty relation" $T\Delta E \gtrsim 1$, where $T, \Delta E$ are the running time and energy resolution of the algorithm, respectively. Thus, a generic (exact) eigenstate cannot be efficiently computed even on a quantum computer, and cannot be efficiently prepared in experiments if any experimental operation allows efficient quantum simulation. In both theory and practice, one could only hope to efficiently obtain an approximate eigenstate with an inverse polynomial energy resolution.

Inputing a microcanonical ensemble of inverse polynomial bandwidth, we propose a (rigorous) polynomial-time (classical) algorithm that (up to negligible errors) outputs a (positive) diagonal density operator (mixed state) supported on the ensemble, expressed as a matrix product operator (MPO). The algorithm first simulates the real-time evolution using the matrix product formalism (this step dominates the running time of the algorithm), and then suppresses unwanted eigenstates using an en-ergy filtering technique [2, 19, 23]. When acting on an eigenstate, the propagator $e^{iHt}$ can be viewed as a (classical) wave whose angular frequency is the eigenenergy. In this language, we design an interference pattern that selects the eigenstates in an energy (frequency) interval of an inverse polynomial bandwidth.

We present a rigorous and a heuristic version of the algorithm. The former should start with a necessary condition for MBL, i.e., violation of this condition implies delocalization. We assume neither (III) nor (IV) because these heuristic characterizations of MBL seem to be quite strong conditions from a rigorous point of view. Rather, we assume that the effect of any local perturbation (e.g., injecting conserved charges) is (up to negligible errors) restricted to a region whose radius grows (at most) logarithmically with time ("logarithmic light cone"). This is a minimum condition: Any system with diffusive or even ballistic propagation of information can be dubbed delocalized. Rewriting this condition as a Lieb-Robinson type inequality [19, 21, 29], we simulate the time evolution efficiently and rigorously using the matrix product formalism. In two and higher spatial dimensions, a modified version of our algorithm outputs a diagonal density operator supported on the microcanonical ensemble, expressed as a projected entangled pair operator (PEPO) [51] of quasi-polynomial bond dimension. However, we do not know how to compute the physical properties of such a PEPO efficiently [46].

The heuristic version of our algorithm is slightly different in that it simulates the time evolution by TEBD, which is the only possibly non-rigorous step (the energy filtering step is always rigorous). Although we do believe that TEBD is indeed rigorous (TEBD never gets stuck in a local minimum because it is not a local search algorithm), we are not aware of such a proof in the literature. This version can be easily coded either from scratch for a DMRG engineer or by using well-established numerical packages [5, 16]. In practice, it can be used to, e.g., detect energy-tuned dynamical quantum phase transitions between MBL phases. Such an unconventional quantum critical point was predicted in the random quantum Ising chain with weak interactions using the excited-state real-space renormalization group (RSRG-X) technique [41]. We are trying to observe this quantum critical point numerically (in preparation).

The quantum phase estimation algorithm is an important subroutine with various applications in quantum computing. It can be used to study MBL [7]. In 1D MBL systems, it can be efficiently simulated by a non-optimal version of our (classical) algorithm.

*Note added.*—A recent article [43] appeared while the present work was being completed. It proposed a heuristic variational approach to approximate the eigenstates of a 1D MBL Hamiltonian based on the assumption that all eigenstates can be mapped to the computational basis states by a single local quantum circuit of small depth. An important open problem is to identify the class of 1D MBL systems in which this assumption is valid. Does it break down in systems with localization-protected topological order [27]?

A fresh article [59] appeared on 18 Aug 2015 after a manuscript of the present work was ready to be but not yet submitted to arXiv. It introduced the spectrum bifurcation renormalization group (SBRG) as a generalization of the real-space renormalization group. Various features of 1D MBL systems (and the quantum phase transitions between them) have been obtained numerically using the SBRG technique. We like this technique very much.

It should be clear that our approach is fundamentally different from those of [43, 59]. Furthermore, the energy resolution of our algorithm is $1/\operatorname{poly} n$, where $n$ is the system size. The energy resolutions of the algorithms in [43, 59] appear to be $c\sqrt{n}$, where $c$ is a small but non-negligible constant. Note that the standard deviation of energy for a generic state is $\Theta(\sqrt{n})$ due to the quantum central limit theorem [11].

*Algorithm.*—The main idea is to annihilate unwanted eigenstates by destructive interference. Suppose we are working with a chain of $n$ spins, and consider a microcanonical ensemble with an energy interval $[E - \epsilon, E + \epsilon]$ whose bandwidth $2\epsilon = 1/\operatorname{poly} n$ is an inverse polynomial in the system size. Let $P$ be the projection onto the subspace spanned by all states in the ensemble. We choose $\{f_j \in \mathbf{C}, t_j \in \mathbf{R}\}$ for $j = 0, 1, \ldots, N - 1$ such that (i) $N = \operatorname{poly} n$; (ii) $|t_j| = \operatorname{poly} n$ such that each $e^{iHt_j}$ can be computed efficiently; (iii) $PQP \approx Q$ such that $Q$ is almost supported on the microcanonical ensemble; (iv)

$$Q = \sum_{j=0}^{N-1} f_j e^{i(H-E)t_j} \qquad (1)$$

such that $Q$ is diagonal in the eigenbasis of $H$. From a technical point of view, we prefer to express $Q$ as an integral

$$Q = \int_{-\infty}^{+\infty} f(t) e^{i(H-E)t} dt, \qquad (2)$$

where $f : \mathbf{R} \to \mathbf{C}$ is called the filter function [2, 19, 23]. We require (i) $\lim_{|t|\to\infty} f(t)$ decays rapidly such that the integral (2) can be truncated to $|t| \le T$ for $T = \operatorname{poly} n$; (ii) $f^*(t) = f(-t)$ such that $Q = Q^\dagger$ is Hermitian.

Filter functions have been systematically constructed. The simplest and most efficient one in the present context is a Gaussian filter:

$$\begin{aligned} Q &:= e^{-\frac{q(H-E)^2}{2\epsilon^2}} = \frac{\epsilon}{\sqrt{2\pi q}} \int_{-\infty}^{+\infty} e^{-\frac{\epsilon^2 t^2}{2q} + i(H-E)t} dt \\ &\approx \frac{\epsilon}{\sqrt{2\pi q}} \int_{-T}^{T} e^{-\frac{\epsilon^2 t^2}{2q} + i(H-E)t} dt \\ &\approx \frac{\epsilon \tau}{\sqrt{2\pi q}} \sum_{j=-T/\tau}^{T/\tau} e^{-\frac{\epsilon^2 j^2 \tau^2}{2q} + i(H-E)j\tau}. \end{aligned} \qquad (3)$$

Let $\|X\|$ and $\|X\|_1 = \text{tr}\sqrt{X^\dagger X}$ denote the operator norm and the trace norm, respectively. $Q$ is almost supported on the microcanonical ensemble in the sense that $\|PQP - Q\| \le e^{-q/2}$. Thus, $\|PQP - Q\|$ is negligible for $q = \text{poly}\, n$. It is easy to see that $\|PQP - Q\|_1$ is also negligible for $q = \text{poly}\, n$. The truncation (the second line in (3)) error is upper bounded by

$$\frac{2\epsilon}{\sqrt{2\pi q}} \int_T^\infty e^{-\frac{\epsilon^2 t^2}{2q}} dt = O\left(e^{-\frac{\epsilon^2 T^2}{2q}}\right), \quad (4)$$

which is negligible for $T = \text{poly}\, n$. The discretization (the third line in (3)) error is polynomially small for $\tau = 1/\text{poly}\, n$.

Heuristically, the propagator $e^{iHt}$ can be computed using the TEBD algorithm [54]. As entanglement grows logarithmically with time in 1D MBL systems [1, 3, 26, 48, 55, 56, 60], we expect that $e^{iHt}$ is well approximated by an MPO of bond dimension $e^{O(\log t)} = \text{poly}\, t$. Assuming that the running time of TEBD is a polynomial in the bond dimension, $e^{iHt}$ can be computed in time $\text{poly}\, t$. Overall, our algorithm outputs an approximation $Q'$ of $Q$ in time $(T/\tau)\text{poly}\, T = \text{poly}\, n$. The correctness of our algorithm can be efficiently verified numerically by checking whether the energy distribution of $Q'$ is sufficiently sharply peaked.

If entanglement grows polylogarithmically with time (e.g., the critical random quantum Ising chain with weak interactions [56]), we expect that $e^{iHt}$ is well approximated by an MPO of bond dimension $e^{\text{poly}\log t}$. In such systems, our algorithm runs in quasi-polynomial time.

The operator $Q$ can be viewed in two complementary ways. First, $Q/\text{tr}\, Q$ is a density operator (mixed state) almost supported on the microcanonical ensemble. Second, $Q$ is an approximate projection onto the ensemble. To prepare a superposition (pure state) of eigenstates in the ensemble, the first heuristic approach is to maximize $\langle\psi|Q|\psi\rangle$ variationally over all MPS $|\psi\rangle$ of small bond dimension. The bond dimension should be increased until the objective function saturates to 1. The second heuristic approach is to start with a random product state $|\psi_0\rangle$, which is trivially an MPS. In the $j$th ($j \ge 1$) step, an MPS $|\psi_j\rangle$ is obtained by compressing the MPS $Q|\psi_{j-1}\rangle$. Repeat these steps until the energy distribution of $|\psi\rangle$ is sufficiently sharply peaked.

To accelerate our algorithm, we can start with a small $q$ such that $Q$ (3) suppresses unwanted eigenstates moderately but not significantly. Then we proceed by repeated squaring. In the $j$th step, we obtain an MPO $Q^{2^j}$. We expect that the bond dimension of $Q^{2^j}$ grows exponentially but not doubly exponentially with $j$ as $Q$ is constructed from a MBL Hamiltonian $H$.

*Proof.*—We now simulate the time evolution efficiently and rigorously. Let $H = \sum_{j=1}^{n-1} H_j$, where $\|H_j\| \le 1$ acts on the spins $j$ and $j+1$. Suppose $H$ is MBL in the sense that the effect of any local perturbation is exponentially suppressed outside a region whose radius grows (at most) logarithmically with time:

$$|\langle\psi|A(t)|\psi\rangle - \langle\psi|U^\dagger A(t)U|\psi\rangle| \le e^{-\Omega(d(A,U))}\text{poly}\, t. \quad (5)$$

Here, $|\psi\rangle$ is an arbitrary initial state; $A(t) := e^{-iHt}Ae^{iHt}$ with $\|A\| \le 1$ is a local observable in the Heisenberg picture; $U$ is a local unitary operator describing the perturbation; $d(A, U)$ is the distance between $A$ and $U$. As any local operator $B$ (with bounded norm) can be expressed as a linear combination of local unitary operators, (5) simplifies to a Lieb-Robinson type inequality

$$\|[A(t), B]\| \le e^{-\Omega(d(A,B))}\text{poly}\, t. \quad (6)$$

Reference [29] proved (without using the matrix product formalism) that (6) implies (i) starting from a product state, the entanglement entropy grows at most logarithmically with time; (ii) the expectation value of a local operator $\langle\hat{O}(t)\rangle$ can be computed in time $\text{poly}(nt/\delta)$ on a classical computer, where $\delta$ is the desired precision.

We now prove a stronger statement: (6) implies that $e^{iHt}$ is approximated by an MPO of bond dimension $\text{poly}(nt/\delta)$; furthermore, such an MPO can be efficiently constructed. Let $I$ be an interval, and define

$$H_I = \sum_{j \in I} H_j, \quad H_k^l(t) = e^{-iH_{(k-l,k+l)}t}H_k e^{iH_{(k-l,k+l)}t}, \quad H_k'(t) = e^{-iH_{(k-2l,n)}t}H_k e^{iH_{(k-2l,n)}t}. \quad (7)$$

(6) implies

$$\|H_k(t) - H_k^l(t)\| = \|\hat{U}^\dagger(t)H_k\hat{U}(t) - H_k\| \le \int_0^t \left\|\frac{d}{d\tau}(\hat{U}^\dagger(\tau)H_k\hat{U}(\tau))\right\| d\tau \le \int_0^t \|[H_k(\tau), H_{(0,k-l]} + H_{[k+l,n)}]\| d\tau$$

$$\le \left(\sum_{j=-\infty}^{k-l} + \sum_{j=k+l}^{+\infty}\right) e^{-\Omega(|j-k|)} \int_0^t \text{poly}\,\tau\, d\tau = e^{-\Omega(l)}\text{poly}\, t, \quad \hat{U}(t) := e^{iHt}e^{-iH_{(k-l,k+l)}t}. \quad (8)$$

Similarly and hence,

$$\|H_k'(t) - H_k^l(t)\| \le \|H_k(t) - H_k^l(t)\| + \|H_k(t) - H_k'(t)\| \le e^{-\Omega(l)}\text{poly}\, t + e^{-\Omega(l)}\text{poly}\, t = e^{-\Omega(l)}\text{poly}\, t. \quad (9)$$

Let $\mathcal{T}$ be the time-ordering operator. Following [38], we consider the unitary operators

$$V_k(t) := e^{-iH_{(k-2l,k)}t}e^{-iH_{(k,n)}t}e^{iH_{(k-2l,n)}t} = \mathcal{T}e^{i\int_0^t H_k'(\tau)d\tau}, \quad V_k^l(t) := \mathcal{T}e^{i\int_0^t H_k^l(\tau)d\tau}. \tag{10}$$

$V_k^l(t)$ acts on the spins $k-l+1, k-l+2, \ldots, k+l$, and is an approximation to $V_k(t)$. Lemma 1 in [25] and (10) imply

$$\|V_k(t) - V_k^l(t)\| \leq e^{-\Omega(l)} \operatorname{poly} t \Rightarrow \left\|e^{iH_{(k-2l,n)}t} - e^{iH_{(k-2l,k)}t}e^{iH_{(k,n)}t}V_k^l(t)\right\| \leq \delta/n \tag{11}$$

for $l = O(\log(tn/\delta))$. Iterating this procedure for $k = 2lk'$ with $k' = 1, 2, \ldots$, we obtain

$$e^{iHt} \approx e^{iH_{(0,2l)}t}e^{iH_{(2l,n)}t}V_{2l}^l(t) \approx e^{iH_{(0,2l)}t}e^{iH_{(2l,4l)}t}e^{iH_{(4l,n)}t}V_{2l}^l(t)V_{4l}^l(t) \approx \cdots \approx \prod_{k'} e^{iH_{(2(k'-1)l,2k'l)}t} \prod_{k'} V_{2k'l}^l(t). \tag{12}$$

The rightmost side is a quantum circuit of depth 2, where each quantum gate is a unitary operator acting on $O(l)$ contiguous spins. Thus, we have explicitly constructed an MPO of bond dimension $e^{O(l)} = \operatorname{poly}(tn/\delta)$ that approximates $e^{iHt}$ to error $\delta$.

*Classical simulation of quantum phase estimation.*— Naively, a quantum computer obtains a random eigenstate by measuring the Hamiltonian. However, this measurement is difficult to perform as it is nonlocal. The standard approach is to estimate the phases of $U := e^{iH\tau}$, where $0 \leq H\tau < 2\pi$ with $\tau = 1/\operatorname{poly} n$ ($n$ is the system size) prevents the eigenvalues of $U$ from wrapping around the unit circle in the complex plane. The propagator $U$ can be constructed using the Trotter decomposition [32] or more advanced methods [8–10].

Let $\{E_k, |\psi_k\rangle\}$ be a complete set of eigenvalues and eigenstates of $H$, and suppose we are given a state $|\psi\rangle = \sum_k c_k|\psi_k\rangle$. In phase estimation, we add an auxiliary register of dimension $T/\tau = \operatorname{poly} n$ (or $\log_2(T/\tau) \in \mathbf{Z}$ ancilla qubits), and label its computational basis by $\{|j\rangle\}_{j=0}^{T/\tau-1}$. We (i) prepare the state $\sum_j |j\rangle$ (a normalization factor is neglected for simplicity); (ii) apply controlled-$U$ gates $T/\tau$ times; (iii) apply the inverse quantum Fourier transform $|j\rangle \to \sum_{J=0}^{T/\tau-1} e^{-2\pi ijJ\tau/T}|J\rangle$; (iv) measure the ancillas:

$$\sum_j |\psi\rangle|j\rangle \to \sum_j U^j|\psi\rangle|j\rangle = \sum_k c_k|\psi_k\rangle \sum_j e^{iE_k j\tau}|j\rangle$$
$$\to \sum_k c_k|\psi_k\rangle \sum_J r(E_k\tau - 2\pi J\tau/T)|J\rangle$$
$$\to \sum_k c_k r(E_k\tau - 2\pi J\tau/T)|\psi_k\rangle|J\rangle, \tag{13}$$

where $r(x) := (e^{ixT/\tau} - 1)/(e^{ix} - 1)$ is peaked at $x = 0$ with (i) $\lim_{x\to 0} r(x) = T/\tau$; (ii) $\max_{c \leq |x| \leq 1} |r(x)| \sim 1/c$ for $\tau/T \ll c \ll 1$; (iii) $|r(x)| \sim 1$ for generic $x$. Thus, the quantum phase estimation algorithm effectively constructs the operator $R := r(H\tau - 2\pi J\tau/T)\tau/T$ for a random $J = 0, 1, \ldots, T/\tau - 1$. Indeed, it fits the framework (1) with a trivial filter function:

$$N = T/\tau, \ E = 2\pi J/T, \ t_j = j\tau, \ f_j = 1, \ \forall j. \tag{14}$$

As such, the quantum phase estimation algorithm can be efficiently simulated classically in 1D MBL systems. The operator $R$ is almost supported on the microcanonical ensemble with energy $[E - \epsilon, E + \epsilon]$ in the sense that $\|PRP - R\| \lesssim 1/(\epsilon T)$, but the trace norm $\|PRP - R\|_1$ cannot be satisfactorily bounded. Thus, a nontrivial filter function can improve the performance of the algorithm.

*Applications.*—We now demonstrate the applications of our algorithm in practical settings. As $Q'$ is an MPO of polynomial bond dimension, its physical properties can be efficiently computed.

The eigenstate thermalization hypothesis (ETH) says that eigenstates at the same energy density have (approximately) the same local expectation values [15, 44, 49]. It is believed that MBL (extended) systems violate (satisfy) ETH. We expect that the local reduced density matrix of $Q/\operatorname{tr} Q$ typically deviates from the identity matrix even in MBL systems. Although a microcanonical ensemble of inverse polynomial bandwidth has an exponential number of eigenstates, the local reduced density matrices of these eigenstates are not completely statistically independent (and do not average to the identity matrix) regardless of whether ETH holds, because the energy density is fixed.

Spectral functions of local operators, which are in principle directly measurable in experiments, can be studied with our method. As the energy resolution of our algorithm is an inverse polynomial in the system size, we can only obtain coarse-grained spectral functions. This already goes beyond experiments as the energy resolution of any apparatus is finite.

Consider the weakly interacting random quantum Ising chain

$$H = \sum_{i \in \mathbf{Z}} J_i \sigma_i^z \sigma_{i+1}^z + h_i \sigma_i^x + J_i' \sigma_i^x \sigma_{i+1}^x, \tag{15}$$

where $J_i$'s are independent and identically distributed (i.i.d.), $h_i$'s are i.i.d., and $J_i'$'s are i.i.d. random variables. We assume $|J_i'|$'s $\ll |J_i|, |h_i|$'s. When $J_i, h_i, J_i' < 0$, RSRG-X predicts a critical energy scale $E_c$ above (below) which (almost) all eigenstates are in the Hilbert-glass (paramagnetic) phase [41]. RSRG-X is (mostly) an analytical approach and involves approximations that are not fully controlled. Hence, it is desirable that the predictions of RSRG-X are checked against ab initio numerical calculations.

This energy-tuned dynamical quantum phase transition is characterized by the spin autocorrelation function. Let $C(t) := \langle\psi|\sigma_i^z(t)\sigma_i^z(0)|\psi\rangle$, where $|\psi\rangle$ is an eigenstate of $H$ with energy $E$. RSRG-X predicts $C(|t| \to +\infty) > 0$ if $E > E_c$ and $C(|t| \to +\infty) = 0$ if $E \leq E_c$ [41]. Furthermore, dynamical RSRG predicts a power-law decay $C(t \gg 1) \sim t^{-2\Delta(E)}$ if $E < E_c$, where $\Delta(E) > 0$ is a monotonically decreasing function of $E$ such that $\Delta(E \to E_c) = 0$ [56]. We expect that $C(t)$ and $C'(t) := \operatorname{tr}(\sigma_i^z(t)\sigma_i^z(0)Q)/\operatorname{tr} Q$ behave similarly if the bandwidth $\epsilon$ is not too large. We are trying to compute $C'(t)$ up to $t = \operatorname{poly} n$ and extract its asymptotic behavior convincingly (in preparation).

*Higher spatial dimensions.*—(3) remains valid in dimensions higher than one. It appears that we should use the projected entangled pair formalism (a generalization of the matrix product formalism to higher dimensions). If entanglement grows polylogarithmically with time, a heuristic counting argument suggests that $e^{iHt}$ and $Q$ are well approximated by PEPO of bond dimensions $e^{\operatorname{poly}\log t}$ and $e^{\operatorname{poly}\log n}$, respectively. In practice, we do not know how to implement TEBD efficiently in higher dimensions. In theory, the rigorous version of our algorithm can be extended to higher dimensions with a little additional effort, and it runs in quasi-polynomial time. However, we do not know how to compute the physical properties of such a PEPO efficiently [46].

The author would like to thank Bela Bauer, Joel E. Moore, and Romain Vasseur for helpful advices. This work was supported by NSF DMR-1206515 and the Institute for Quantum Information and Matter at the California Institute of Technology.

---